\documentstyle[preprint,aps,epsf]{revtex}
\tightenlines
\begin{document}

\title{Color Current Induced by Gluon in Background Field Method of QCD}

\author{Qun Wang$^{1,2}$, Chung-Wen Kao$^{1}$, 
Gouranga C. Nayak$^{1}$, Horst St\"ocker$^{1}$ 
and Walter Greiner$^{1}$}

\address
{$^1$\small\it{Institut f\"ur Theoretische Physik,
J. W. Goethe-Universit\"at,
60054 Frankfurt am Main, Germany}
$^2$\small\it{Physics Department, 
Shandong University, 
Jinan, Shandong 250100, P. R. China} 
}

\date{\today} 

\maketitle

\begin{abstract}
By using the background field method of QCD 
in a path integral approach, 
we derive the equation of motion for the 
classical chromofield and for the gluon 
in a system containing the gluon and the classical chromofield 
simultaneously. This inhomogeneous 
field equation contains a current term, 
which is the expectation value of a composite 
operator including linear, square and cubic terms 
of the gluon field. We also derive 
identities which the current should obey from the gauge 
invariance. We calculate the current 
at the leading order where 
the current induced by the gluon is opposite in sign 
to that induced by the quark. This is just the feature 
of the non-Abelian gauge field theory 
which has asymptotic freedom. 
Physically, the induced current 
can be treated as the 'displacement' current 
in the polarized vacuum, and its effect is equivalent to 
redefining the field and the coupling constant. 
\end{abstract}


\section{Introduction}

The behavior of electric charged particles in a classical
electromagnetic field has been extensively 
studied in the literature. However, the behavior of color charged 
particles (quarks and gluons) in a classical chromofield has 
been less studied. The gluon case is particularly interesting because 
there is no analogue in the electromagnetic theory, 
since photons do not interact with the 
classical electromagnetic field. 
One important issue is to derive the equation of motion 
for the gluon in the presence of chromofield and to define 
and calculate the classical current produced by the gluon. 
As we know it is easy to define and understand 
the color current for quarks in the classical chromofield. 
But strictly defining the color current 
for gluons classically is not an easy task, mainly 
due to its quantum and bosonic nature. 
We plan to address some aspects 
of these issues in this paper.

At first sight, this problem looks as if it were only 
of theoretical interest. However, it turns out to be applicable to 
a non-equilibrium system of quarks 
and gluons, which is expected to be produced 
in ultra-relativistic heavy ion collisions 
in RHIC and LHC where a classical chromofield 
may be produced. The dynamics of a quantum 
non-equilibrium many-body system 
is usually studied in the closed-time-path formulation 
with non-local source terms\cite{ctp}. 
This formalism needs to be 
extended by incorporating classical chromofield. 
The evolution of partons can be obtained by solving 
the Schwinger-Dyson equation defined on a closed-time-path. 
This is formidable task because it involves a non-local 
non-linear integrodifferential equation and because of 
its quantum nature. There are two scales 
in the system: the quantum (microscopic) scale and 
the statistical-kinetic (macroscopic) one. 
When the statistical-kinetic scale is much larger 
than the quantum one, the Schwinger-Dyson equation 
may be recasted into a much simpler form of 
the kinetic Boltzmann equation by 
a gradient expansion. The Boltzmann equation describes 
evolution of the particle distribution function 
in the phase space of momentum-coordinate, and can be solved 
numerically for practical purpose. 
For a non-equilibrium system of quarks 
and gluons in a chromofield, the distribution 
function also depends on the classical 
color charge of the parton, since 
the color is exchanged between the chromofield 
and partons and among partons themselves. 
In this case, the Boltzmann equation also describes 
the evolution of the parton distribution function 
in the color space\cite{heinz,wong}. 
While the Boltzmann equation describes 
the kinetic and color evolution of the 
hard parton system, soft partons are normally 
treated as a coherent classical field whose evolution
may be described by a equation which is 
similar to Yang-Mills equation.
Therefore, one should study the transport 
problem for hard partons in the 
presence of a classical background field. 
For example, in high energy heavy-ion collisions, 
minijets (which are hard partons) are initially produced 
and then propagate in a classical chromofield 
created by the soft partons\cite{qgp,bhal,larry}. 
In this situation it is necessary to derive the 
equation of the gluon and the classical 
chromofield to study the formation and 
equilibration of quark-gluon plasma. 
Hence we see that the issues we want to address 
in this paper have practical implication 
in ultra-relativistic heavy ion collisions.

Since conventional QCD does not contain 
classical chromofield, to include the classical chromofield 
in a proper way, we follow the background field 
method of QCD (BG-QCD). This method was first introduced 
by DeWitt and 't Hooft\cite{dewitt,thooft}. 
The advantage of BG-QCD is that it is formulated 
in an explicit gauge 
invariance manner(see below). This method is widely used 
to study confinement, gravity and hot 
quark-gluon-plasma, etc.. In this paper, we derive 
the equation of motion for the gluon and 
the classical background field to define the induced current 
in a path integral approach which fully 
incorporate non-Abelian gauge symmetry. 
Also some properties of the induced current are 
given in the framework of BG-QCD.

The paper is organized as follows. 
In section II, we briefly introduce basics of 
BG-QCD and the Ward identity which we will 
use later. In section III, equations of motion for 
the classical and the gluon field are derived, 
and the current induced by the gluon is defined. 
We study its property and derive the Ward 
identity that the current obeys in the next section. 
Section V presents the result for the 
current at the leading order. The physical implication 
for the result is also discussed. 
The summary and conclusion is presented 
in the last section. 

\section{Background Field Method}

We know that any physical quantity calculated 
in the conventional QCD is gauge 
invariant and independent of 
the particular gauge chosen. 
However, there is still a problem of 
gauge invariance which is evident in the 
off shell Green's functions and counter terms. 
This problem arises because one must fix a gauge 
and introduce the gauge-fixing and ghost 
term to quantize the theory. 
The classical Lagrangian for the gauge field has explicit
gauge is invariant while the gauge-fixing and ghost term
is not gauge invariant although it is invariant 
under the BRS transformation.
The background field method is such a technique
which allows one to fix a gauge (background-field gauge)
without losing explicit gauge invariance which is 
present in the classical level of the gauge field theory.

By writing the conventional gluon field into 
the sum of a classical background part and the quantum one, 
the action in the background field 
gauge is given by\cite{abb81,soh86,zub75,lee73,elze90}: 
\begin{equation}
\begin{array}{l}
S=S_0+S_{fix}+S_{ghost}+S_{source} ,\\
S_0=-\frac 14 \int d^4x (F_{\mu \nu}^i[A]+D_{\mu}^{ij}[A]Q_{\nu}^j-
D_{\nu}^{ij}[A]Q_{\mu}^j+gf^{ijk}Q_{\mu}^jQ_{\nu}^k)^2 ,\\
S_{fix}=-\frac {1}{2\alpha}\int d^4x (D_{\mu}^{ij}[A]Q^{\mu ,j})^2 ,\\
S_{ghost}=\int d^4x \overline{C}^i D_{\mu}^{ij}[A]
D^{\mu ,jk}[A+Q]C^k ,\\
S_{source}=\int d^4x (J_{\mu}^i Q^{\mu ,i}
+\overline{\xi}^i C^i +\overline{C}^i \xi ^i).
\end{array}
\end{equation}
The generating functional for Green's function is
\begin{equation}
\label{Z}
Z[A,J,\xi ,\overline{\xi}]=\int [dQ][dC][d\overline{C}] \exp (iS).
\end{equation}
The variables are defined as follows: 
$A_{\mu}^i$ is the classical background field; 
$Q_{\mu}^i$ is the gluon field, 
the variable of integration in the functional integral; 
$C^i, \overline{C}^i$ are ghost and anti-ghost fields; 
$\overline{\xi}^i, \xi ^i$ are external sources 
coupling to ghost and anti-ghost respectively. 
$D_{\mu}^{ij}[A]$ is the covariant derivative in
the adjoint representation whose generators
are $(t_A^a)^{ij}=if^{iaj}$ and it is defined as
$D_{\mu}^{ij}[A]\equiv \partial _{\mu}\delta ^{ij}
+gf^{iaj}A_{\mu}^a$=
$\partial _{\mu}\delta ^{ij}-ig(t_A^{a})^{ij}A_{\mu}^a$; 
$F_{\mu \nu}^i[A]$ is the field strength 
tensor for the background field and given 
by $F_{\mu \nu}^i[A]=\partial _{\mu}A_{\nu}^i
-\partial _{\nu}A_{\mu}^i+gf^{ijk}A_{\mu}^jA_{\nu}^k$.

The gauge transformation for all fields and sources
which leaves $Z$ invariant is given by:
\begin{equation}
\label{t1}
\begin{array}{l}
A'_{\mu}=UA_{\mu}U^{-1}+\frac ig U\partial _{\mu}U^{-1},\\
Q'_{\mu}=UQ_{\mu}U^{-1},\,\,\, J'_{\mu}=UJ_{\mu}U^{-1},\\
C'_{\mu}=UCU^{-1},\,\,\, \overline{\xi}'=U\overline{\xi}U^{-1},\\
\overline{C}'=U\overline{C}U^{-1},\,\,\, \xi '=U\xi U^{-1}.
\end{array}
\end{equation}
This gauge transformation is known as Type-I 
transformation\cite{zub75}. 
Note that the background field transforms 
in the conventional way while the gluon field, 
the ghost field and all external sources transform like 
a matter field. The infinitesimal changes for 
these fields and sources are: 
\begin{equation}
\label{s-tr}
\begin{array}{l}
\delta A_{\mu}^i=D_{\mu}^{ij}[A]\omega ^j,\\
\delta Q_{\mu}^i=gf^{ijk}Q_{\mu}^j \omega ^k,\,\,\,\, 
\delta J_{\mu}^i=gf^{ijk}J_{\mu}^j \omega ^k ,\\
\delta C^i=gf^{ijk}C^j \omega ^k,\,\,\,\,
\delta \overline{\xi}^i=gf^{ijk}\overline{\xi}^j \omega ^k ,\\
\delta \overline{C}^i=gf^{ijk}\overline{C}^j \omega ^k,\,\,\,
\delta \xi ^i=gf^{ijk}\xi ^j \omega ^k .\\
\end{array}
\end{equation}
Here we define the gauge transformation as 
$U(x)=\exp (ig\omega ^a(x)t_A^a)$. Roman 
alphabets denote color indices and Greek ones denote 
the Lorentz indices. All fields and sources 
are either in the vector form, e.g. 
$A_{\mu}\equiv A_{\mu}^at_A^a$, or with their color indices
explicitly shown, e.g. $A_{\mu}^a$. It is obvious that 
the generating functional $Z$ is invariant 
under the above transformation. 
But the less obvious is the invariance of $S_{fix}$. 
To prove it, one should 
remember the fact that $t_A^a$ is a pure imaginary 
Hermitian matrix and that $(t_A^a)^T=-t_A^a$, 
therefore $U^T=U^{-1}$.

The generating functional for the connected Green function 
is given by:
\begin{equation}
W[A,J,\xi,\overline{\xi}]=-i\ln Z[A,J,\xi,\overline{\xi}]
\end{equation}
Field averages can be obtained by taking derivative of 
$W$ with respect to their corresponding sources:
\begin{equation}
\langle Q_{\mu}^i\rangle =\frac{\delta W}{\delta J^{\mu,i}},\,\,\,\,
\langle C^i\rangle =\frac{\delta W}{\delta \overline{\xi}^i},\,\,\,\,
\langle \overline{C}^i\rangle =-\frac{\delta W}{\delta \xi ^i} 
\end{equation}
The one-particle irreducible generating functional 
$\Gamma [A,\langle Q\rangle ,\langle C\rangle ,
\langle \overline{C}\rangle ]$ is defined by Legendre 
transformation for $W$:
\begin{equation}
\Gamma [A,\langle Q\rangle ,\langle C\rangle ,
\langle \overline{C}\rangle ]=W[A,J,\xi,\overline{\xi}]
-\int d^4x (J_{\mu}^i \langle Q^{\mu ,i}\rangle 
+\overline{\xi}^i\langle C^i\rangle 
+\langle \overline{C}^i\rangle \xi ^i)
\end{equation}
where we see that $\frac{\delta \Gamma}{\delta A_{\mu}^j}
=\frac{\delta W}{\delta A_{\mu}^j}$. 
Since $\Gamma [A,\langle Q\rangle ,\langle C\rangle ,
\langle \overline{C}\rangle ]$ is a gauge invariant 
functional of $A$, $\langle Q\rangle$, $\langle C\rangle$,
and $\langle \overline{C}\rangle$, hence 
$\Gamma [A,\langle Q\rangle ,\langle C\rangle ,
\langle \overline{C}\rangle ]$ should be invariant under 
the transformation (\ref{s-tr}). Then we have 
the following Ward identity\cite{zub75}: 
\begin{equation}
\label{WI1}
D_{\mu}^{ij}[A]\frac{\delta \Gamma}{\delta A_{\mu}^j}+
gf^{ijk}(\langle Q_{\mu}^j\rangle
\frac{\delta \Gamma}{\delta \langle Q_{\mu}^k\rangle }+
\langle C^j\rangle 
\frac{\delta \Gamma}{\delta \langle C^k\rangle }+
\langle \overline{C}^j\rangle 
\frac{\delta \Gamma}{\delta \langle \overline{C}^k\rangle })=0
\end{equation}
In the same way, we have the parallel Ward identity 
in terms of $W$:
\begin{equation}
\label{WI2}
D_{\mu}^{ij}[A]\frac{\delta W}{\delta A_{\mu}^j}+
gf^{ijk}(J_{\mu}^j\frac{\delta W}{\delta J_{\mu}^k}+
\xi ^j\frac{\delta W}{\delta \xi ^k}+
\overline{\xi}^j\frac{\delta W}{\delta \overline{\xi}^k})=0.
\end{equation}
Note that Eq.(\ref{WI1}) and (\ref{WI2}) are 
identical because $\frac{\delta \Gamma}{\delta A_{\mu}^j}
=\frac{\delta W}{\delta A_{\mu}^j}$. 

\section{Equation of motion for gluon and classical field and 
Definition of induced current}

By shifting the quantum field $Q_{\mu}^i(x)$, 
a variable of integration in the
functional integral, by a small amount $v_{\mu}^i(x)$: 
$Q'^i_{\mu}(x)=Q_{\mu}^i(x)+v_{\mu}^i(x)$, 
we can obtain the equation of motion or 
the Lagrange equation for the quantum field $Q_{\mu}^i$. 
The change of the functional for connected 
Green's functions $W$ is: 
\begin{equation}
\delta W=-i\delta \ln Z=Z^{-1}
\int [dQ][dC][d\overline{C}]
\int d^{4}x
\frac{\delta S}{\delta Q_{\mu}^i}v_{\mu}^i(x)\exp (iS).
\end{equation}
The fact that $W$ should be independent of the shifting value 
of the integration variable leads to the equation of motion 
for $Q_{\mu}^i$: 
\begin{equation}
\langle\frac{\delta S}{\delta Q_{\mu}^i(x)}\rangle=
Z^{-1}\int [dQ][dC][d\overline{C}]
\left(\frac{\delta (S_0+S_{fix}+S_{ghost})}{\delta Q_{\mu}^i(x)}
+J^{\mu ,i}(x)\right)\exp (iS)=0.
\end{equation}
In the absence of the external source $J_{\mu}^i$, 
the equation of motion becomes: 
\begin{equation}
\langle\frac{\delta S_0}{\delta Q_{\mu}^i}\rangle+
\langle\frac{\delta S_{fix}}{\delta Q_{\mu}^i}\rangle+
\langle\frac{\delta S_{ghost}}{\delta Q_{\mu}^i}\rangle=0, 
\end{equation}
where we have 
\begin{equation}
\label{em1}
\begin{array}{l}
\frac{\delta S_0}{\delta Q_{\mu}^i}
=D_{\nu}^{ij}[A+Q]F^{\nu\mu,j}[A+Q],\\ 
\frac{\delta S_{fix}}{\delta Q_{\mu}^i}
=\frac{1}{\alpha}D^{\mu,ij}[A]D^{\nu,jk}[A]Q_{\nu}^k ,\\
\frac{\delta S_{ghost}}{\delta Q_{\mu}^i}
=-gf^{kil}\overline{C}^j
\stackrel{\leftarrow}{D}\raisebox{0.5ex}{$^{\mu,kj}$}[A]C^l.
\end{array}
\end{equation}
Note that the partial derivative operation in 
$\stackrel{\leftarrow}{D}\raisebox{0.5ex}{$^{\mu,kj}$}$ 
is toward $\overline{C}^j$. 
We know that $F^{\nu\mu,j}[A+Q]$ transforms 
like a matter field and $D_{\nu}^{ij}[A+Q]$ is commutable with 
the transformation matrix. So the first equation 
$\frac{\delta S_0}{\delta Q_{\mu}^i}
=D_{\nu}^{ij}[A+Q]F^{\nu\mu,j}[A+Q]$ 
transforms like a matter field in adjoint 
representation. Similarly we can easily 
verify that the second equation transforms 
like a matter field. But it is less apparent 
to verify that the ghost part 
$\frac{\delta S_{ghost}}{\delta Q_{\mu}^i}$ 
also transforms like a matter field. Let us verify it 
directly. The variation of 
$\frac{\delta S_{ghost}}{\delta Q_{\mu}^i}$ 
under the infinitesimal gauge transformation 
Eq.(\ref{s-tr}) is:
\begin{equation}
\begin{array}{rll}
\delta (\frac{\delta S_{ghost}}{\delta Q_{\mu}^i})
&=&-gf^{kil}\delta (\overline{C}^j
\stackrel{\leftarrow}{D}\raisebox{0.5ex}{$^{\mu,kj}$}[A])C^l
-gf^{kil}\overline{C}^j
\stackrel{\leftarrow}{D}\raisebox{0.5ex}{$^{\mu,kj}$}[A]\delta C^l \\
&=&-gf^{aid}(gf^{ake}\overline{C}^j
\stackrel{\leftarrow}{D}\raisebox{0.5ex}{$^{\mu,kj}$}[A]\omega ^e)C^d
-gf^{kil}\overline{C}^j
\stackrel{\leftarrow}{D}\raisebox{0.5ex}{$^{\mu,kj}$}[A]
(gf^{lde}C^d\omega ^e)\\
&=&-g^2(f^{aid}f^{ake}+f^{kil}f^{lde})\overline{C}^j
\stackrel{\leftarrow}{D}\raisebox{0.5ex}{$^{\mu,kj}$}[A]C^d\omega ^e\\
&=&gf^{ile}[-gf^{kld}\overline{C}^j
\stackrel{\leftarrow}{D}\raisebox{0.5ex}{$^{\mu,kj}$}[A]C^d]\omega ^e
\end{array}
\end{equation}
Therefore we see that $\frac{\delta S_{ghost}}{\delta Q_{\mu}^i}$ 
transforms like a matter field in adjoint representation. 

We can rewrite Eq.(\ref{em1}) in another form:
\begin{equation}
\label{eqm}
D_{\nu}^{ij}[A]F^{\nu\mu,j}[A]
=\langle j_0^{\mu ,i}\rangle
\end{equation}
where the current $j_0$ induced by the quantum field is given by: 
\begin{equation}
\label{jq}
\begin{array}{lll}
j_0^{\mu ,i}&=&
-gf^{abc}D_{\nu}^{ia}[A](Q^{\nu, b}Q^{\mu ,c}) 
-gf^{ida}Q_{\nu}^d(D^{\nu ,ab}[A]Q^{\mu, b}-D^{\mu ,ab}[A]Q^{\nu, b})\\
&&-g^2f^{ida}f^{abc}Q_{\nu}^{d}Q^{\nu ,b}Q^{\mu ,c}
-D_{\nu}^{ij}[A](D^{\nu ,jk}[A]Q^{\mu, k}-D^{\mu ,jk}[A]Q^{\nu, k})
-gf^{ijk}Q_{\nu}^jF^{\nu\mu,k}[A]\\
&&+gf^{kil}\overline{C}^j
\stackrel{\leftarrow}{D}\raisebox{0.5ex}{$^{\mu,kj}$}[A]C^l
-\frac{1}{\alpha}D^{\mu,ij}[A]D^{\nu,jk}[A]Q_{\nu}^k
\end{array}
\end{equation}
The last two terms in $j_0$ are 
just $-\delta (S_{fix}+S_{ghost})/\delta Q$. 
Obviously $j_0$ transforms like a matter field. 

It is easy to check $\delta S_0/\delta Q=\delta S_0/\delta A$.
But it takes some steps to prove:
\begin{equation}
\label{sq-sa}
\langle\frac{\delta (S_{fix}+S_{ghost})}{\delta Q_{\mu}^i}\rangle
=\langle\frac{\delta (S_{fix}+S_{ghost})}{\delta A_{\mu}^i}\rangle .
\end{equation}
We start with $Z[A,0,0,0]$ where in Eq.(\ref{Z}) 
the sources are set to zero. We shift 
$A$ by $\delta A$ and $Q$ by $-\delta A$ so that 
$A+Q$ does not change. $S_0$ is invariant 
in this operation. The variation of $Z[A,0,0,0]$ is 
caused by that of $(S_{fix}+S_{ghost})$:
\begin{equation}
\label{dz1}
\begin{array}{rl}
\delta Z[A,0,0,0]=&i\int [dQ][dC][d\overline{C}] 
\{\frac{\delta (S_{fix}+S_{ghost})}{\delta A_{\mu}^i}
-\frac{\delta (S_{fix}+S_{ghost})}{\delta Q_{\mu}^i}
\}\delta A_{\mu}^i\exp (iS)\\
=&i\int [dQ][dC][d\overline{C}]
\{-\frac{1}{\alpha}D^{\mu,ij}[A+Q]D^{\nu,jk}[A]Q_{\nu}^k\\
&-gf^{ijk}\overline{C}^jD^{\mu,kl}[A+Q]C^l\}\delta A_{\mu}^i\exp (iS)
\end{array}
\end{equation}
where $S=S_0+S_{fix}+S_{ghost}$. 
We expect that it would leave $Z[A,0,0,0]$ unchanged 
i.e. $\delta Z[A,0,0,0]/\delta A_{\mu}^i=0$, since
nothing happens for $S_0[A+Q]$. But 
it is non-trivial to prove this. We provide here 
a proof based on Ref.\cite{zub75} and \cite{lee73}. 
Consider a kind of special infinitesimal 
gauge transformation defined by: 
$\delta Q_{\mu}^i=D_{\mu}^{ij}[A+Q]\omega ^j$ 
where the infinitesimal 
gauge group parameter $\omega ^j(x)$ is defined by:
\begin{equation}
\omega ^j(x)=M^{-1,jk}(x,y)D_{\nu}^{kl}[A+Q](y)
\delta A_{\nu}^l(y) 
\end{equation}
where the repetition of indices implies the summation 
or the integration; The matrix $M$ is just the one 
which couples to the ghost field: 
\begin{equation}
M^{ij}(x,y)=D_{\mu}^{ik}[A]D_{\nu}^{kj}[A+Q](x)\delta (x-y)
\end{equation}
we have following properties:
\begin{equation}
D_{\mu}^{ij}[A](x)\delta Q_{\mu}^j(x)=M^{ik}(x,y)\omega ^k(y)
=D_{\mu}^{ij}[A+Q](x)\delta A_{\mu}^j(x)
\end{equation}
We say this infinitesimal gauge transformation is 
special because it can make gauge condition invariant in some 
circumstances. In this paper, we imply the covariant 
gauge constraint $\delta (D_{\mu}^{ij}[A]Q^{\mu,j}-a^i)$ in 
the generating functional $Z$. If we increase $a^i$ by a small 
amount $\delta a^i$ and simultaneously make the above 
gauge transformation where the small parameter $\omega ^k(y)$ 
is given the value so that $M^{ik}(x,y)\omega ^k(y)=\delta a^i$, 
the gauge constraint remains the same. In fact the new 
gauge constraint 
\begin{equation}
\begin{array}{rll}
\delta (D_{\mu}^{ij}[A]Q'^{\mu,j}-a'^i)
&=&\delta (D_{\mu}^{ij}[A]Q^{\mu,j}
+D_{\mu}^{ij}[A]\delta Q_{\mu}^j-a^i-\delta a^i)\\
&=&\delta (D_{\mu}^{ij}[A]Q^{\mu,j}-a^i)
\end{array}
\end{equation}
is unchanged. The other part of 
the generating functional $Z$ can be proved to be 
invariant under the above gauge transformation constrained by 
$M^{ik}(x,y)\omega ^k(y)=\delta a^i$. Hence $Z$ is 
independent of $a^i$. One can therefore multiplies an 
arbitrary function of $a^i$ and integrate over $a^i$ 
without changing the result up to some irrelevant 
normalization constant. Usually this function of $a^i$ 
is chosen to be a Gaussian function 
$\exp (-i\int d^4x(a^{i}(x))^2/2\alpha )$ which gives 
the gauge fixing term $S_{fix}$. 
One can prove that the variation $\delta Z$ 
caused by the above defined gauge transformation constrained by 
$M^{ik}(x,y)\omega ^k(y)=\delta a^i$ is just $\delta Z[A,0,0,0]$ 
in Eq.(\ref{dz1}) which is zero. 
Therefore we obtain 
$\langle\delta (S_{fix}+S_{ghost})/\delta A\rangle
=\langle\delta (S_{fix}+S_{ghost})/\delta Q\rangle$. 
Together with 
$\delta S_0/\delta Q=\delta S_0/\delta A$, we have 
\begin{equation}
\langle\delta (S_0+S_{fix}+S_{ghost})/\delta A\rangle
=\langle\delta (S_0+S_{fix}+S_{ghost})/\delta Q\rangle=0
\end{equation}

From $\langle\delta (S_0+S_{fix}+S_{ghost})/\delta A\rangle=0$, 
we obtain: 
\begin{equation}
\label{eqm1}
D_{\nu}^{ij}[A]F^{\nu\mu,j}[A]=\langle j^{\mu ,i}\rangle ,
\end{equation}
where $j$ is defined as: 
\begin{equation}
\label{ja}
\begin{array}{rll}
j^{\mu,i}&=&-gf^{abc}D_{\nu}^{ia}[A](Q^{\nu, b}Q^{\mu ,c}) 
-gf^{ida}Q_{\nu}^d(D^{\nu ,ab}[A]Q^{\mu, b}-D^{\mu ,ab}[A]Q^{\nu, b})\\
&&-g^2f^{ida}f^{abc}Q_{\nu}^{d}Q^{\nu ,b}Q^{\mu ,c}
-D_{\nu}^{ij}[A](D^{\nu ,jk}[A]Q^{\mu, k}-D^{\mu ,jk}[A]Q^{\nu, k})
-gf^{ijk}Q_{\nu}^jF^{\nu\mu,k}[A]\\
&&-gf^{kil}\overline{C}^j
\stackrel{\leftarrow}{D}\raisebox{0.5ex}{$^{\mu,kj}$}[A]C^l
-gf^{ijk}\overline{C}^jD^{\mu,kl}[A+Q]C^l
+\frac 1{\alpha}gf^{ijk}Q^{\mu,k}D_{\nu}^{jl}[A]Q^{\nu,l} 
\end{array}
\end{equation}
where the last three terms are just 
$\delta (S_{fix}+S_{ghost})/\delta A_{\mu}^i$. 
Because of the identity 
$D_{\mu}^{ij}[A]D_{\nu}^{jk}[A]F^{\nu\mu,k}[A]=0$, 
we have $D_{\mu}^{ij}[A]\langle j^{\mu ,j}\rangle=0$. 

Eq.(\ref{eqm}), Eq.(\ref{sq-sa}) and Eq.(\ref{eqm1}) 
are our main results. They guarantee 
the consistent definition of the induced current. 
As we see it is easier and clearer to prove 
the equation of motion for $\langle Q\rangle$, 
$\langle\frac{\delta S}{\delta Q}\rangle =0$. 
But it is less apparent for 
$\langle\frac{\delta S}{\delta A}\rangle =0$, 
the equation of motion for the classical background field, in a 
quantum field formalism. As a contrast, it is 
straightforward to obtain the Lagrange equation of motion 
for a field in the classical field theory, but the current 
situation is subtle because the 
treatment of the quantum field, 
the gauge fixing term as well
as the ghost field and is highly non-trivial. 
The proof for the equivalence of two equations 
is guaranteed by the intrinsic relation between the 
the gauge fixing term and the ghost sector, 
or in other words, the equivalence and the consistency of 
two equations is a natural outcome of the symmetry 
behind the non-Abelian gauge field theory. 
We notice that Ref.\cite{elze90} also derives 
Eq.(\ref{eqm1}) from a canonical formalism, 
but what we use here is a path integral approach 
where the symmetry of non-Abelian gauge field is 
explicitly fulfilled and exploited.

\section{Identity For Induced Current}

We now try to derive the equation for the current 
from the Ward identity. 
In order to derive the equation for the current 
from the Ward identity, 
we add to $S_{source}$ a new 
source term $\int d^4x Y_{\mu}^i j^{\mu ,i}$ where  
the current $j$ induced by the 
quantum field couples to its 
external source $Y$. Since $j$ transforms like 
a matter field, we require that 
$Y$ also transform in the same way 
to ensure the gauge invariance of 
the new source term. 

Before we proceed, some points in the 
derivation should be clarified.
We distinguish two cases: 

{\bf Case I.} There is no constraint on 
values of $\langle Q_{\mu}^i\rangle=\delta W/\delta J^{\mu,i}$, 
$\langle C^i\rangle=\delta W/\delta \overline{\xi ^i}$ and 
$\langle\overline{C}^i\rangle=-\delta W/\delta \xi ^i$ in 
intermediate steps. Therefore, 
during the derivation there is no 
relation between external sources 
($J$, $\xi$, $\overline{\xi}$) and 
the background field $A$. Hence sources and 
the background field are independent variables. 
Because we do not impose any specific 
value on $\langle j\rangle$ in the whole process, the source 
$Y$ remains irrelevant of $A$. 

{\bf Case II.} $\langle Q\rangle$, $\langle C\rangle$ and 
$\langle\overline{C}\rangle$ are set to zero in each step,
as assumed by many other authors\cite{abb81,soh86}, 
hence specific relations between 
sources ($J$, $\xi$, $\overline{\xi}$)
and $A$ are built up by 
$\delta W[J,\xi,\overline{\xi}]/\delta J^{\mu,i}=0$, 
$\delta W/\delta \overline{\xi ^i}=0$ and 
$\delta W/\delta \xi ^i=0$. 
Each source is a functional of 
$A$ and vice versa. Through $A$, 
there are functional mappings between any two sources too. 
These sources are no longer independent of each other. 
But we do not assume any specific 
value for $\langle j\rangle$, so
the source $Y$ is still independent of $A$. 
In this case, we have two independent 
variables, $Y$ and $A$. 

So in both cases $Y$ is always an independent variable. 
We will see in either case, 
i.e. whether $\langle Q\rangle,\langle C\rangle ,
\langle\overline{C}\rangle=0$ or not, 
the equation for $\langle j\rangle$ is the same, though 
the Ward identity for each case is different. 
Here we assume case I(For case II, see Appendix A). 
After introducing a new source term, 
the Ward identity (\ref{WI2}) is extended to: 
\begin{equation}
\label{WI3}
D_{\mu}^{ij}[A]\frac{\delta W}{\delta A_{\mu}^j}+
gf^{ijk}(Y_{\mu}^j\frac{\delta W}{\delta Y_{\mu}^k}+
J_{\mu}^j\frac{\delta W}{\delta J_{\mu}^k}+
\xi ^j\frac{\delta W}{\delta \xi ^k}+
\overline{\xi}^j\frac{\delta W}{\delta \overline{\xi}^k})=0,
\end{equation}
where $\delta W/\delta A_{\mu}^j$ is given by:
\begin{equation}
\label{dwda}
\begin{array}{lll}
\frac{\delta W}{\delta A_{\mu}^j}
&=&\langle\frac{\delta (S_0+S_{fix}+S_{ghost})}{\delta A_{\mu}^j}\rangle
+Y_{\nu}^i\langle\frac{\delta j^{\nu,i}}{\delta A_{\mu}^j}\rangle\\
&=&D_{\nu}^{jk}[A]F^{\nu\mu,k}[A]-\langle j^{\mu ,j}\rangle 
+Y_{\nu}^k\langle\frac{\delta j^{\nu,k}}{\delta A_{\mu}^j}\rangle .
\end{array}
\end{equation}
Substituting Eq.(\ref{dwda}) into Eq.(\ref{WI3}) 
and using the identity 
$D_{\mu}^{ij}[A]D_{\nu}^{jk}[A]F^{\nu\mu,k}[A]=0$, 
Eq.(\ref{WI3}) becomes:
\begin{equation}
\label{WI4}
-D_{\mu}^{ij}[A]\langle j^{\mu ,j}\rangle 
+D_{\mu}^{ij}[A]Y_{\nu}^k
\langle\frac{\delta j^{\nu,k}}{\delta A_{\mu}^j}\rangle
+gf^{ijk}(Y_{\mu}^j\frac{\delta W}{\delta Y_{\mu}^k}+
J_{\mu}^j\frac{\delta W}{\delta J_{\mu}^k}+
\xi ^j\frac{\delta W}{\delta \xi ^k}+
\overline{\xi}^j\frac{\delta W}{\delta \overline{\xi}^k})=0.
\end{equation}
Because we want to go further, we must keep all 
sources in Eq.(\ref{WI4}). If we set all 
sources vanish at this stage, we get 
$D_{\mu}^{jk}[A]\langle j^{\mu,k}\rangle=0$ which 
we obtain in the former section. 
Taking derivative with respect to $Y_{\nu}^k(y)$ 
for Eq.(\ref{WI4}), setting all sources vanish, 
and using $D_{\mu}^{jk}[A]\langle j^{\mu ,k}\rangle=0$, 
we obtain:
\begin{equation}
\label{eqj1}
\begin{array}{l}
D_{\mu}^{ij}[A](x)\langle T[j^{\mu ,j}(x)j^{\nu ,k}(y)]\rangle
+i\delta ^4(x-y)gf^{ilj}A_{\mu}^l(x)
\langle\frac{\delta j^{\nu,k}}{\delta A_{\mu}^j}(x)\rangle\\
+ig\delta ^4(x-y)f^{ikl}\langle j^{\nu,l}(x)\rangle
+i\partial _{\mu}[\delta ^4(x-y)\langle\frac{\delta j^{\nu,k}}
{\delta A_{\mu}^i}(x)\rangle]=0.
\end{array}
\end{equation}
Integrating Eq.(\ref{eqj1}) over $x$, 
and noting that the complete differential vanishes, we obtain:
\begin{equation}
\label{eqj2}
f^{ilj}\int d^4xA_{\mu}^{l}(x)
\langle T[j^{\mu ,j}(x)j^{\nu ,k}(y)]\rangle
+if^{ilj}A_{\mu}^l(y)
\langle\frac{\delta j^{\nu,k}}{\delta A_{\mu}^j}(y)\rangle
+if^{ikl}\langle j^{\nu,l}(y)\rangle=0.
\end{equation}
In addition, integrating 
$D_{\mu}^{jk}[A]\langle j^{\mu ,k}(x)\rangle=0$ over 
$x$, we obtain: 
\begin{equation}
\label{eqj3}
\int d^4x f^{ilk}A^l_{\mu}(x)\langle j^{\mu ,k}(x)\rangle =0
\end{equation}
Eq.(\ref{eqj1}-\ref{eqj3}) are identities for $j$. 
In above formula, we use the language of 
functional path integral and the expectation 
value is taken in a functional sense. 
We can translate path integral expressions into 
the language of quantum operator due to 
the exact correspondence between two approaches. 
Then the expectation operation is carried on 
the interacting vacuum state, i.e. that of the full 
Hamiltonian $H=H_0+H_i$ where $H_i$ denotes the 
interacting Hamiltonian. In above formula, 
all operators are in the Heisenberg picture. According to 
Gell-Mann-Low's theorem, we can rewrite them in the 
interacting picture:
$\langle (\cdots )_H \rangle
\equiv \langle \Omega |(\cdots )_H |\Omega \rangle
=\langle 0|(\cdots )_I e^{-i\int dtH_{iI}(t)}|0\rangle$ 
where $|\Omega \rangle$ is the vacuum state for the 
full Hamiltonian $H$ and $|0\rangle$ is that for 
the free $H_0$, and 'H' and 'I' denote the 
Heisenberg and interacting picture respectively. 

Note that $\frac{\delta j}{\delta A}(x)$ and 
$j(x)$ are composite operators, so they should be 
understood as normal product in order to be well defined in 
taking their expectation values on the vacuum state. 
If we assume, $|\Omega\rangle \approx |0\rangle $, 
Eq.(\ref{eqj1}) is then:
\begin{equation}
D_{\mu}^{ij}[A](x)\langle 0|T[j^{\mu ,j}(x)j^{\nu ,k}(y)]|0 \rangle=0.
\end{equation}
where we drop in $j$ linear terms of $Q$; Also we have used 
$\langle 0|\frac{\delta j^{\nu,k}}{\delta A_{\mu}^j}(x)|0\rangle=0$ 
and $\langle 0|j^{\nu,l}(x)|0\rangle=0$ because 
$\frac{\delta j^{\nu,k}}{\delta A_{\mu}^j}(x)$ and 
$j^{\nu,l}(x)$ are treated as normal products. 

Because we notice that $\frac{\delta j}{\delta A}$ 
transforms in a good manner, similarly we can also 
add to the action a new source term, where 
$\frac{\delta j}{\delta A}$ couples 
to its external source, to obtain a new identity. 
For simplicity, we drop linear terms 
in $j$ which results in that 
$\frac{\delta j}{\delta A}$ is independent of $A$. 
Defining $\left(\frac{\delta j}{\delta A}\right)^{\mu\nu,ab}
\stackrel{\rm def}{=}\frac{\delta j^{\mu,a}}{\delta A_{\nu}^b}$, 
we can check the transformation rule: 
$\left(\frac{\delta j}{\delta A}\right)'^{\mu\nu,ab}
=(U\left(\frac{\delta j}{\delta A}\right)^{\mu\nu}U^{-1})^{ab}$. 
We add to the action a new source term 
$\tilde{Y}_{\mu\nu}^{ba}\left(\frac{\delta j}{\delta A}\right)^{\mu\nu,ab}$. 
To make the source term invariant under the gauge transformation, 
we assume that $\tilde{Y}$ transforms in the same way as 
$(\frac{\delta j}{\delta A})$ does, i.e. $\tilde{Y}'^{ba}_{\mu\nu}
=(U\tilde{Y}_{\mu\nu}U^{-1})^{ba}$. The infinitesimal variation of 
$\tilde{Y}$ is then: 
\begin{equation}
\delta \tilde{Y}_{\mu\nu}^{ba}
=-g\omega ^i(f^{bim}\tilde{Y}_{\mu\nu}^{ma}
-f^{nia}\tilde{Y}_{\mu\nu}^{bn})
\end{equation}
The Ward identity becomes:
\begin{equation}
\begin{array}{l}
D_{\mu}^{ij}[A]\frac{\delta W}{\delta A_{\mu}^j}
+g(f^{bim}\tilde{Y}_{\mu\nu}^{ma}
-f^{nia}\tilde{Y}_{\mu\nu}^{bn})
\frac{\delta W}{\delta \tilde{Y}_{\mu\nu}^{ba}}\\
+gf^{ijk}(J_{\mu}^j\frac{\delta W}{\delta J_{\mu}^k}+
\xi ^j\frac{\delta W}{\delta \xi ^k}+
\overline{\xi}^j\frac{\delta W}{\delta \overline{\xi}^k})=0
\end{array}
\end{equation}
Because $\frac{\delta j}{\delta A}$ 
is independent of $A$, only 
$\frac{\delta S}{\delta A}$ finally contributes 
to $\frac{\delta W}{\delta A}$. Repeating the same 
procedure as in the derivation of 
Eq.(\ref{eqj1}) and (\ref{eqj2}), we can obtain:
\begin{equation}
\label{eqj4}
\begin{array}{l}
D_{\mu}^{ij}[A](x)\langle T[j^{\mu ,j}(x)
\frac{\delta j^{\beta,l}}{\delta A_{\lambda}^k}(y)]\rangle\\
+ig\delta ^4(x-y)(f^{bil}
\langle\frac{\delta j^{\beta,k}}{\delta A_{\lambda}^b}(x)\rangle
+f^{kia}\langle\frac{\delta j^{\beta,a}}
{\delta A_{\lambda}^l}(x)\rangle)=0.
\end{array}
\end{equation}
The integration form is: 
\begin{equation}
gf^{ilj}\int d^4xA_{\mu}^{l}(x)
\langle T[j^{\mu ,j}(x)
\frac{\delta j^{\beta,l}}{\delta A_{\lambda}^k}(y)]\rangle
+ig(f^{bil}\langle\frac{\delta j^{\beta,k}}{\delta A_{\lambda}^b}(x)\rangle
+f^{kia}\langle\frac{\delta j^{\beta,a}}
{\delta A_{\lambda}^l}(y)\rangle)=0 .
\end{equation}

We can take derivative with respect to $Y$ for 
Eq.(\ref{WI4}) arbitrary times and get a series of identities, 
each of them shows a relation among 
correlation functions for $j$ of the lower order 
and those of the higher order. This is similar to 
the BBKGY hierarchy in statistical physics\cite{balescu}. 

\section{Current at leading order}

In this section, as an example, we calculate the 
expectation value of the current 
$\langle j^{\mu ,i}(x)\rangle$ at 
the leading order. Since $j^{\mu ,i}(x)$ is a 
composite operator, in order to be well defined, 
it is treated as normal product. 
If the expectation value $j^{\mu ,i}(x)$ 
is taken between the vacuum state 
of a free field Lagrange, we simply 
have $\langle 0|j^{\mu ,i}(x)|0\rangle=0$. 
So $\langle j^{\mu ,i}(x)\rangle$ starts 
from the loop level. Here we consider 
the lowest order result at the one-loop level. 
At this level, the relevant term in the action 
is $-\int d^4xj^i_{\mu}A^{\mu,i}$. The corresponding 
term in $H_i$ is then $\int d^3xj^i_{\mu}A^{\mu,i}$, 
so $-i\int dtH_i=-i\int d^4xj^i_{\mu}A^{\mu,i}$. 
Hence we obtain: 
 \begin{equation}
\label{j-lo}
\langle j^{\mu ,i}(x)\rangle
=-i\int d^4y A_{\nu}^j(y)
\langle 0|T[j_{A^0}^{\mu ,i}(x)j_{A^0}^{\nu ,j}(y)]|0\rangle 
\end{equation}
where $j_{A^0}$ is part of 
$j$ in Eq.(\ref{ja}), which is independent of $A$ 
and has only square terms of the $Q$ field; 
$-\int d^4y A_{\nu}^j(y)j_{A^0}^{\nu ,j}(y)$ is 
part of the classical action which contributes 
to the lowest order. $j_{A^0}$ is given by: 
\begin{equation}
\begin{array}{rll}
j_{A^0}^{\mu ,a}&=&
-gf^{abc}\partial _{\nu}(Q^{\nu,b}Q^{\mu,c})
+gf^{abc}Q^{\nu,c}\partial _{\nu}Q^{\mu,b}
-gf^{abc}Q^{\nu,c}\partial ^{\mu}Q_{\nu}^b\\
&&-\frac 1{\alpha}gf^{abc}Q^{\mu,c}\partial _{\nu}Q^{\nu,b}
-gf^{abc}\overline{C}^b
\stackrel{\leftarrow}{\partial}\raisebox{0.5ex}{$^{\mu}$}C^c
+gf^{abc}\overline{C}^b
\partial ^{\mu}C^c  
\end{array}
\end{equation}
where we have not included terms 
proportional to $Q$, $QQQ$ and $Q\overline{C}C$. 
Because there are only square and cubic terms 
of $Q$ in $j$, linear terms of $Q$ in the action 
do not contribute to the result. 
We drop the $QQQ$ term both in $j$ and 
in the classical action because it 
contributes to the two-loop level. 

Note that Eq.(\ref{j-lo}) should satisfy Eq.(\ref{eqj3}): 
\begin{equation}
\label{eqj6}
\int d^4x f^{ilk}A^l_{\mu}(x)\langle j^{\mu ,k}(x)\rangle 
=-if^{ilk}\int d^4xd^4y A^l_{\mu}(x)A_{\nu}^j(y)
\langle 0|T[j_{A^0}^{\mu ,k}(x)j_{A^0}^{\nu ,j}(y)]|0\rangle 
=0
\end{equation}
Actually, for the leading-order $j$ given in Eq.(\ref{j-lo}), 
Eq.(\ref{eqj2}) and Eq.(\ref{eqj3}) are equivalent. 
Noticing in Eq.(\ref{eqj3}) the term $if^{ilj}A_{\mu}^l(y)
\langle\frac{\delta j^{\nu,k}}{\delta A_{\mu}^j}(y)\rangle $ 
gives an additional strong coupling constant $g$ 
relative to rest terms, we can neglect it 
for the leading-order $j$ and obtain: 
\begin{equation}
\label{eqj5}
f^{ilj}\int d^4xA_{\mu}^{l}(x)
\langle T[j_{A^0}^{\mu ,j}(x)j_{A^0}^{\nu ,k}(y)]\rangle
+if^{ikl}\langle j^{\nu,l}(y)\rangle=0
\end{equation} 
where only $j_{A^0}$ contributes to the first term 
and other parts of $j$, i.e. $j_{A^1}$ and $j_{A^2}$, 
contribute to the higher order. 
Multiplying the above equation by $A_{\nu}^k(y)$ and 
integrating it over $y$, we obtain:
\begin{equation}
\label{eqj7}
f^{ilk}\int d^4xd^4yA_{\mu}^{l}(x)A_{\mu}^{m}(y)
\langle T[j_{A^0}^{\mu ,k}(x)j_{A^0}^{\nu ,m}(y)]\rangle
+if^{ilk}\int d^4yA_{\nu}^l(y)\langle j^{\nu,k}(y)\rangle=0
\end{equation}
Due to Eq.(\ref{eqj6}), the above equation is actually:
\begin{equation}
2if^{ilk}\int d^4yA_{\nu}^l(y)\langle j^{\nu,k}(y)\rangle=0
\end{equation} 
Surely it is the same as Eq.(\ref{eqj6}) itself.

The self-energy tensor $\Pi ^{\mu\nu,ab}(x,y)
=\langle0|T[j_{A^0}^{\mu ,a}(x)j_{A^0}^
{\nu ,b}(y)]|0\rangle$ 
corresponds to one-loop diagrams of the (quantum-)gluon and the 
ghost with two external $A$ legs(see fig.1). In momentum space, 
$\Pi ^{\mu\nu,ab}(p)$ is given by: 
\begin{equation}
\begin{array}{l}
\Pi ^{\mu\nu,ab}(p)=\Pi _Q^{\mu\nu,ab}(p)+\Pi _C^{\mu\nu,ab}(p),\\
\Pi _Q^{\mu\nu,ab}(p)=ig^2C_A\delta ^{ab}
(g^{\mu\nu}p^2-p^{\mu}p^{\nu})\frac 1{(4\pi )^2}\frac {10}{3} 
(\frac 1{\epsilon _r}+\ln (4\pi )-\gamma+\frac {59}{30}
+\ln (\frac{\mu^2}{-p^2-i\epsilon })),\\
\Pi _C^{\mu\nu,ab}(p)=ig^2C_A\delta ^{ab}
(g^{\mu\nu}p^2-p^{\mu}p^{\nu})\frac 1{(4\pi )^2}\frac {1}{3} 
(\frac 1{\epsilon _r}+\ln (4\pi )-\gamma+\frac{8}{3}
+\ln (\frac{\mu^2}{-p^2-i\epsilon })),
\end{array}
\end{equation}
where $\Pi _Q^{\mu\nu,ab}$ is from the gluon loop 
and $\Pi _C^{\mu\nu,ab}$ from the ghost loop; 
$C_A=N_c$ for $SU(N_c)$; $\mu ^2$ is 
a mass scale; $\gamma$ is the Euler's constant; 
The pole term $1/\epsilon _r=2-D/2$ arises 
in the regularization. 
We calculate $\Pi _Q^{\mu\nu,ab}$ with the gauge 
parameter $\alpha =1$, i.e. we choose the Feynman gauge. 
The term $\frac 1{\epsilon _r}+\ln (4\pi )-\gamma$ is to be 
cancelled by the counterterm in $\overline{MS}$ scheme. 
After the reduction of the divergent part for $\Pi ^{\mu\nu,ab}$, 
we obtain: 
\begin{equation}
\label{n-munu}
\Pi ^{\mu\nu,ab}(p)=ig^2\frac 1{(4\pi )^2}\frac {11}{3}C_A
\delta ^{ab}(g^{\mu\nu}p^2-p^{\mu}p^{\nu})
\ln (\frac{\mu^2}{-p^2-i\epsilon }).
\end{equation}
Then we take the Fourier transformation and get 
$\Pi ^{\mu\nu,ab}(x,y)$:
\begin{equation}
\label{pixy}
\begin{array}{rll}
\Pi ^{\mu\nu,ab}(x,y)&=&\int \frac{d^4p}{(2\pi )^4}
\Pi ^{\mu\nu,ab}(p)\exp (-ip(x-y))\\ 
&=&-ig^2C_A\frac 1{(4\pi )^2}\frac {11}{3} 
\delta ^{ab}(g^{\mu\nu}\partial ^2-\partial ^{\mu}\partial ^{\nu}) H(x-y).
\end{array}
\end{equation} 
where $\partial $ acts on $x$ or $y$ and 
$H(x)$ is the Fourier transformation of 
$\ln (\frac{\mu^2}{-p^2-i\epsilon })$ 
and its derivation is given in Appendix B. 
Substituting Eq.(\ref{pixy}) into 
Eq.(\ref{j-lo}), we derive: 
\begin{equation}
\label{j-1l}
\langle j^{\mu ,a}(x)\rangle=-g^2C_A\frac 1{(4\pi )^2}\frac {11}{3}
\int d^4y H(x-y)
(g^{\mu\nu}\partial ^2-\partial ^{\mu}\partial ^{\nu})A_{\nu}^a(y)
\end{equation}
where $\partial $ acts on $A(y)$. 

It is interesting to compare Eq.(\ref{n-munu}) 
with the self-energy tensor of the quark 
loop with two external $A$ legs: 
\begin{equation}
\label{n-munu-q}
\Pi ^{\mu\nu,ab}_{q-loop}=-ig^2\frac 1{(4\pi )^2}\frac {2}{3}n_f
\delta ^{ab}(g^{\mu\nu}p^2-p^{\mu}p^{\nu})
\ln (\frac{\mu^2}{-p^2-i\epsilon })
\end{equation} 
Note that Eq.(\ref{n-munu}) and (\ref{n-munu-q}) 
have opposite signs. This is just the feature of 
the non-Abelian gauge field which is of the same 
origin as the asymptotic freedom and the confinement. 
It means that the current generated by the gluons 
is opposite in sign to that by the quark. 
In Eq.(\ref{n-munu}), we require $p^2<0$ to 
give a real current $j(x)$. If $p^2>0$, 
the current $j(x)$ has an imaginary part which 
would be related to the production rate for the 
real gluon pair. 

In fact, in the derivation of $\Pi ^{\mu\nu,ab}(p)$, 
there is a small positive number $i\epsilon$ 
in the gluon and ghost propagator. 
It finally appears in $\ln \frac{\mu ^2}{-p^2-i\epsilon }$. 
Then we can get the real 
part of $\Pi ^{\mu\nu,ab}(p)$ which 
corresponds to the production rate by noting 
$Im\ln \frac{\mu ^2}{-p^2-i\epsilon }=-\pi$. 
Similar to the definition of 
the production rate for $q\overline{q}$ in \cite{zuber}, 
we obtain the production rate for $gg$: 
\begin{equation}
\label{r-qq}
\begin{array}{rll}
R_{gg}&=&Re\int d^4pA_{\mu}^a(p)\Pi ^{\mu\nu,ab}(p)A_{\nu}^b(-p)\\
&=&\frac 14\frac {11}{3}C_A\alpha _s 
\int d^4p \theta (p^2)
[p^2(A^a(p)\cdot A^a(-p))-(p\cdot A^a(p))(p\cdot A^a(-p))]
\end{array}
\end{equation}
Comparing this rate with that for 
producing $q\overline{q}$ in the classical field: 
\begin{equation}
\label{r-gg}
R_{q\overline{q}}=-\frac 14\frac {2}{3}n_f\alpha _s 
\int d^4p \theta (p^2)
[p^2(A^a(p)\cdot A^a(-p))-(p\cdot A^a(p))(p\cdot A^a(-p))]
\end{equation}
Note that $R_{q\overline{q}}$ has the same form and sign as 
$R_{e^+e^-}$, the rate for the electron-positron pair 
production in QED, while $R_{gg}$ has opposite sign. 
We remember that if $A_{cl}\rightarrow gg$ and 
$A_{cl}\rightarrow q\overline{q}$ are real physical processes, 
$R_{gg}$ and $R_{q\overline{q}}$ 
can be written as $R_{gg}=|M(A_{cl}\rightarrow gg)|^2$ 
and $R_{q\overline{q}}=|M(A_{cl}\rightarrow q\overline{q})|^2$ 
respectively, therefore we have $R_{gg},R_{q\overline{q}}\geq 0$, 
but from the above result Eq.(\ref{r-qq},\ref{r-gg}) 
they have opposite signs. 
Hence, the leading order term in $g$ does not 
give the correct result for the $gg$ pair 
production. We note that the leading order (in $g$) 
gluon pair production term is not gauge invariant with 
respect to the gauge transformation given by 
Eq.(\ref{t1}). In order to obtain correct 
gauge invariant result for gluon 
pair production we have to add higher order terms in $g$, 
which contains three and four $A$'s, into this result. 
For details about this topic, see \cite{dennis}.

Now we try to understand the physical implication of the 
current induced by the gluon. 
In the lowest order, Eq.(\ref{eqm1}) can be written as: 
\begin{equation}
\label{eqm2}
D_{\nu}^{ij}[A(x)]F^{\nu\mu,j}[A(x)]=
-g^2C_A\frac 1{(4\pi )^2}\frac {11}{3}\int d^4y H(x-y)
(g^{\mu\nu}\partial ^2-\partial ^{\mu}\partial ^{\nu})A_{\nu}^a(y)
\end{equation}
In the lowest order, this equation of 
motion for $A$ corresponds to 
the effective action:
\begin{equation}
\begin{array}{rll}
S_{eff}[A]&=&-\int d^4x\frac 14F_{\mu\nu}^j[A(x)]F^{\mu\nu,j}[A(x)]\\
&&+g^2C_A\frac 1{(4\pi )^2}\frac {11}{6}\int d^4xd^4y A_{\mu}^a(x)H(x-y)
(g^{\mu\nu}\partial ^2-\partial ^{\mu}\partial ^{\nu})A_{\nu}^a(y)
\end{array}
\end{equation} 
We can write $S_{eff}[A]$ in momentum space as: 
\begin{equation}
\begin{array}{rll}
S_{eff}[A]&=&-\frac 12\int d^4pA^a_{\mu}(-p)A^a_{\nu}(p)
(g^{\mu\nu}p^2-p^{\mu}p^{\nu})\\
&&\cdot [1+g^2C_A\frac 1{(4\pi )^2}\frac {11}{3}
\ln (\frac{\mu^2}{-p^2-i\epsilon })]\\
&&+{\rm square\ and\ cubic\ terms\ of\ }A(p)\\
\end{array}
\end{equation}
One can define a new amplitude $A'(p)$ with 
the radiation correction caused by 
the gluon in the lowest 
order: $A'(p)=[1+\kappa (p)]^{1/2}A(p)$ 
where $\kappa (p)=g^2\frac 1{(4\pi )^2}\frac {11}{3}C_A
\ln(\frac{\mu^2}{-p^2-i\epsilon })$. 
Therefore, the correction effect brought 
by the gluon fluctuation 
on the classical background field is similar to the medium 
screening effect. This screening effect becomes larger 
as $|p^2|$ decreases and lower as $|p^2|$ increases. 
This is just the feature of the effective coupling 
constant. In fact, since there is a relation 
between the renormalization constant $Z_A$ 
for $A$ and $Z_g$ for $g$: $Z_g=Z_A^{-1/2}$, 
$gA$ should be unchanged after redefinition of 
the field and the coupling constant. 
Now we redefine the field as 
$A'(p)=[1+\kappa (p)]^{1/2}A(p)$, accordingly we can redefine 
the coupling constant as 
$g^{2}_{eff}(p)=[1+\kappa (p)]^{-1}g^{2}(p)$. 
The above argument is made for 
the gluon situation. If we include the quark 
sector into the classical action, 
we should add a term to $\kappa (p)$: 
$\kappa (p)=g^2\frac 1{(4\pi )^2}[\frac {11}{3}C_A-\frac 23n_f]
\ln(\frac{\mu^2}{-p^2-i\epsilon })$. 
It shows the screening effect brought by the quark is 
opposite to that by the gluon. 
We see that the effective coupling constant $g^2_{eff}(p)$ 
has a running property and shows the 
feature of asymptotic freedom. Expressed by the new field and the 
effective coupling constant $g_{eff}$, we can write the original 
field equation (\ref{eqm2}) as: 
$D_{\nu}^{ij}[A',g_{eff}]F^{\nu\mu,j}[A',g_{eff}]=0$. 
Therefore, the current induced by the quantum 
gluon is just the `displacement' current in the polarized 
vacuum, and its effect is equivalent to redefinition of 
the field and the coupling constant.

\section{Summary}

In summary, for a system which contain a classical 
chromofield and the gluon simultaneously, 
we derive the equation of motion for 
the classical chromofield and for the gluon. 
Furthermore, we prove that 
the equation for the classical field is 
the same as for the gluon in a functional 
path integral sense. The proof 
is not straightforward, it needs a 
symmetry property between the gauge 
fixing and ghost sectors of the full action. 
The inhomogeneous field equation 
contains a current term which is 
the expectation value of a composite 
operator including linear, square and cubic terms 
of the quantum field $Q$. We call the current as 
induced by the gluon. 
Note that this current essentially 
is classical quantity. We also derive 
the identity for this current. These identities represent
the constraint on this current from the gauge invariance.
We calculate the current at the leading order. 
In momentum space, 
it is just the contraction of the classical 
field $A(p)$ with the self energy 
tensor $\Pi (p)$ with two $A$'s. 
The current generated by the gluon 
is opposite in sign to that generated by the quark. 
This is just the feature of 
the non-Abelian gauge field theory which is asymptotic free. 
We require $p^2<0$ to give a real current $j(x)$. If $p^2>0$, 
the current $j(x)$ has an imaginary part which 
would give the production rate for the real gluon pair. 
Physically, the current induced by the quantum 
gluon is just the 'displacement' current in the polarized 
vacuum, and its effect is equivalent to redefinition of 
the field and the coupling constant. 

We emphasize that all the analysis done here is with respect 
to QCD vacuum where no medium effect is included yet. 
Normally one works in closed-time path integral 
formalism in order to derive a current 
in a medium. Then the current can be related 
to the non-equilibrium distribution function of the gluon.  
We plan to address this issue in the future where we wish to combine
closed-time path integral formalism with background field
method of QCD. 

\acknowledgements 
Q.W., C.W.K. and G.C.N. acknowledge the financial support 
from Alexander von Humboldt Foundation. 
Q.W. is also supported in part by the 
National Natural Science Foundation of China. 

\newpage

\begin{figure}
\setlength{\epsfxsize}{4.5in}
\centerline{\epsffile{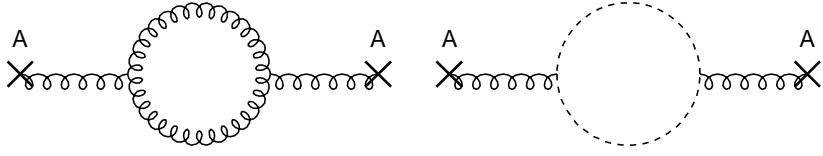}}
\vspace*{.5in}
\caption{
The one-loop diagrams for the self-energy 
with two external $A$ legs. One diagram is the gluon 
loop(screw line) and the other for the ghost loop(dashed line). }
\vspace*{.4in}
\end{figure}

\bigskip

{\bf Appendix A.}

In case II, since all sources are functionals of two 
independent variables $A$ and $Y$, 
the Ward identity (\ref{WI3}) changes to: 
\begin{equation}
\label{WI5}
D_{\mu}^{ij}[A]\frac{\delta W}{\delta A_{\mu}^j}+
gf^{ijk}Y_{\mu}^j\frac{\delta W}{\delta Y_{\mu}^k}=0
\end{equation}
where $\delta W/\delta A_{\mu}^j$ and 
$\delta W/\delta Y_{\mu}^k$ are given by: 
\begin{equation}
\label{dwda1}
\begin{array}{rl}
\frac{\delta W}{\delta A_{\mu}^j}
=&D_{\nu}^{jk}[A]F^{\nu\mu,k}[A]-\langle j^{\mu ,j}\rangle 
+Y_{\nu}^k\langle\frac{\delta j^{\nu,k}}{\delta A_{\mu}^j}\rangle\\
&+\frac{\delta J_{\nu}^k}{\delta A_{\mu}^j}\langle Q^{\nu,k}\rangle
+\frac{\delta \overline{\xi}^k}{\delta A_{\mu}^j}\langle C^k\rangle
+\langle\overline{C}^k\rangle\frac{\delta \xi ^k}{\delta A_{\mu}^j}\\


\frac{\delta W}{\delta Y_{\mu}^k}
=&\langle j^{\mu ,k}\rangle 
+\frac{\delta J_{\nu}^m}{\delta Y_{\mu}^k}\langle Q^{\nu,m}\rangle
+\frac{\delta \overline{\xi}^m}{\delta Y_{\mu}^k}\langle C^m\rangle
+\langle \overline{C}^m\rangle\frac{\delta \xi ^m}{\delta Y_{\mu}^k}
\end{array}
\end{equation}
Since all the field averages are set to zero during our derivation, 
$\frac{\delta W}{\delta A_{\mu}^j}$ is the same as Eq.(\ref{dwda}) 
and $\frac{\delta W}{\delta Y_{\mu}^k}=\langle j^{\mu ,k}\rangle $. 
So we also get Eq.(\ref{eqj2}).

{\bf Appendix B.}

To derive the Fourier transform of $\ln[-p^2-i\epsilon]$, we set:
\begin{equation}
I(x,\xi)\equiv\int\frac{d^{4}p}{(2\pi)^4}e^{-ip\cdot x}\ln[-p_{0}^{2}+\xi^{2}
|\vec{p}|^2-i\epsilon].
\end{equation}
To calculate $I(x,\xi)$, we first evaluate 
the following function $J(x,\xi)$:
\begin{equation}
2\xi J(x,\xi)\equiv\frac{dI(x,\xi)}{d\xi}
=\int\frac{d^{4}p}{(2\pi)^4}e^{-ip\cdot x}
\frac{2\xi|\vec{p}|^2}{-p_{0}^{2}+\xi^{2}
|\vec{p}|^2-i\epsilon}
\end{equation}
First integrate over $p_{0}$:
\begin{eqnarray}
J(x,\xi)&=&-\int\frac{d^{3}p}{(2\pi)^3}\int\frac{dp_{0}}{(2\pi)}
\frac{e^{-ip_{0}x^{0}+i\vec{p}\cdot\vec{x}}|\vec{p}|^2}
{(p_{0}-\xi|\vec{p}|+i\epsilon)(p_{0}+\xi|\vec{p}|-i\epsilon)} \nonumber\\
&=&i\theta(x^{0})\int\frac{d^{3}p}{(2\pi)^3}
\frac{e^{-i\xi|\vec{p}|x^0+i\vec{p}\cdot\vec{x}}}
{2\xi|\vec{p}|}|\vec{p}|^2+i\theta(-x^{0})\int\frac{d^{3}p}{(2\pi)^3}
\frac{e^{i\xi|\vec{p}|x^0+i\vec{p}\cdot\vec{x}}}
{2\xi|\vec{p}|}|\vec{p}|^2 
\end{eqnarray}
Then integrate angular variables:
\begin{eqnarray}
2\xi J(x,\xi)&=&\theta(x^{0})\int_{0}^{\infty}\frac{p^{2}dp}{(2\pi)^2}
\frac{e^{-i\xi px^{0}}}{|\vec{x}|}
(e^{ip|\vec{x}|}-e^{-ip|\vec{x}|})\nonumber\\
&&+\theta(-x^{0})\int_{0}^{\infty}\frac{p^{2}dp}{(2\pi)^2}
\frac{e^{i\xi px^{0}}}{|\vec{x}|}
(e^{ip|\vec{x}|}-e^{-ip|\vec{x}|})
\end{eqnarray}
Next, integrate with $\xi$:
\begin{eqnarray}
I(x,\xi)&=&\frac{\theta(x^{0})}{|\vec{x}|}
\int_{0}^{\infty}\frac{p^{2}dp}{(2\pi)^2}
\frac{e^{-i\xi px^{0}}}{-ipx^{0}}
(e^{ip|\vec{x}|}-e^{-ip|\vec{x}|})\nonumber\\
&&+\frac{\theta(-x^{0})}{|\vec{x}|}
\int_{0}^{\infty}\frac{p^{2}dp}{(2\pi)^2}
\frac{e^{i\xi px^{0}}}{ipx^{0}}
(e^{ip|\vec{x}|}-e^{-ip|\vec{x}|}) \nonumber\\
&=&-\frac{\theta(x^{0})}{|\vec{x}|(x^{0})^{2}}\frac{d}{d\xi}\int_{0}^{\infty}
\frac{dp}{(2\pi)^{2}}e^{-i\xi px^{0}}
(e^{ip|\vec{x}|}-e^{-ip|\vec{x}|})\nonumber\\
&&-\frac{\theta(-x^{0})}{|\vec{x}|(x^{0})^{2}}
\frac{d}{d\xi}\int_{0}^{\infty}
\frac{dp}{(2\pi)^{2}}e^{i\xi px^{0}}
(e^{ip|\vec{x}|}-e^{-ip|\vec{x}|}) \nonumber \\
&=&\frac{-i}{2\pi^{2}}[\frac{\theta(x^{0})}{|\vec{x}|(x^{0})^{2}}
\cdot\frac{4\xi (x^{0})^{2}|\vec{x}|}
{[|\vec{x}|^{2}-\xi^{2}(x^{0})^{2}+i\epsilon ]^{2}}\nonumber\\
&&+\frac{\theta(-x^{0})}{|\vec{x}|(x^{0})^{2}}
\cdot\frac{4\xi (x^{0})^{2}|\vec{x}|}
{[|\vec{x}|^{2}-\xi^{2}(x^{0})^{2}-i\epsilon]^{2}}]
\end{eqnarray}
The final result becomes:
\begin{equation}
I(x,\xi)=
\frac{-i\xi}{\pi^{2}}\left[\frac{\theta(x_{0})}
{[|\vec{x}|^{2}-\xi^{2}(x^{0})^{2}+i\epsilon]^{2}}
+\frac{\theta(-x_{0})}
{[|\vec{x}|^{2}-\xi^{2}(x^{0})^{2}-i\epsilon]^{2}}\right]
\end{equation}
Taking $\xi$=1, we get: 
\begin{eqnarray}
&&\int\frac{d^{4}p}{(2\pi)^4}e^{-ip\cdot x}\ln[-p_{0}^{2}+
|\vec{p}|^2-i\epsilon]\nonumber\\
&&=\frac{-i}{\pi^{2}}\left[\frac{\theta(x_{0})}
{[|\vec{x}|^{2}-(x^{0})^{2}+i\epsilon]^{2}}
+\frac{\theta(-x_{0})}
{[|\vec{x}|^{2}-(x^{0})^{2}-i\epsilon]^{2}}\right]
\end{eqnarray}
Then $H(x)$ is given by:
\begin{eqnarray}
H(x)&=&\int\frac{d^{4}p}{(2\pi)^4}e^{-ip\cdot x}\ln[\frac{\mu ^2}{-p^{2}
-i\epsilon}]=\delta (x)\ln \mu ^2 \nonumber\\
&&-\frac{i}{\pi^{2}}\left[\frac{\theta(x_{0})}
{[|\vec{x}|^{2}-(x^{0})^{2}+i\epsilon]^{2}}
+\frac{\theta(-x_{0})}
{[|\vec{x}|^{2}-(x^{0})^{2}-i\epsilon]^{2}}\right]
\end{eqnarray}

\end{document}